\documentclass{article}

\usepackage{arxiv}

\usepackage[utf8]{inputenc} % allow utf-8 input
\usepackage[T1]{fontenc}    % use 8-bit T1 fonts
\usepackage{hyperref}       % hyperlinks
\usepackage{url}            % simple URL typesetting
\usepackage{booktabs}       % professional-quality tables
\usepackage{amsfonts}       % blackboard math symbols
\usepackage{nicefrac}       % compact symbols for 1/2, etc.
\usepackage{microtype}      % microtypography
\usepackage{lipsum}		% Can be removed after putting your text content
\usepackage{graphicx}
\usepackage[backend=biber, sorting=none]{biblatex} % biblatex with biber
\usepackage{doi}
\usepackage{subfig}

\title{Exploring the Limits of Data Augmentation For Retinal Vessel Segmentation}

\date{} 			

\author{ E. Sadi Uysal  \\
	Department of Computer Engineering\\
	Yildiz Technical University\\
	İstanbul, Turkey \\
	\texttt{enessadi@gmail.com} \\
	\And
	M. Şafak Bilici  \\
	Department of Computer Engineering\\
	Yildiz Technical University\\
	İstanbul, Turkey \\
	\texttt{safakk.bilici@gmail.com} \\
	\And
  B. Selin Zaza \\
  Department of Industrial Engineering\\
  Yildiz Technical University\\
  İstanbul, Turkey \\
  \texttt{b.selinzaza@gmail.com} \\  
  \And
  M. Yiğit Özgenç \\
  Department of Quantitative Methods\\
  Marmara University\\
  İstanbul, Turkey \\
  \texttt{mehmetyigitozgenc@gmail.com} \\
  \And
  Onur Boyar \thanks{Corresponding author.}\\
  Department of Computational Science and Engineering\\
  Boğaziçi University\\
  İstanbul, Turkey \\
  \texttt{boyaronur@gmail.com } \\
}

\hypersetup{
pdftitle={A template for the arxiv style},
pdfsubject={cs.CV},
pdfauthor={E. Sadi Uysal, M. Safak Bilici, B. Selin Zaza, M. Yiğit Özgenç, Onur Boyar},
}
\makeatletter
\newcommand\footnoteref[1]{\protected@xdef\@thefnmark{\ref{#1}}\@footnotemark}
\makeatother
\begin{document}
\maketitle

\begin{abstract}
	Retinal Vessel Segmentation is important for the diagnosis of various diseases. The research on retinal vessel segmentation focuses mainly on the improvement of the segmentation model which is usually based on U-Net [1] architecture. In our study, we use the U-Net architecture and we rely on heavy data augmentation in order to achieve better performance. The success of the data augmentation relies on successfully addressing the problem of input images. By analyzing input images and performing the augmentation accordingly we show that the performance of the U-Net model can be increased dramatically. Results are reported using the most widely used retina dataset, DRIVE.
\end{abstract}

% keywords can be removed

\section{Introduction}

Medical images come in different forms, including  X-ray, computed tomography (CT), magnetic resonance imaging (MRI), ultrasound imaging, fundoscopic images and more. The medical experts analyze the targeted area in  these kinds of images to make a diagnosis. This process can be made much easier by performing segmentation to regions that are meaningful for the task at hand. The segmentation task is achieved by extracting the target class while ignoring the objects from other classes. There are various approaches to solve these segmentation problems. There are signal processing-based approaches \cite{1677727}, heuristic techniques \cite{Nguyen2013AnER}, Support Vector Machine based applications \cite{4336179}, and there are also deep learning applications \cite{10.1007/978-3-319-46723-8_16, kamran2021rvgan}, to name a few. Deep Learning applications have been becoming more popular in recent years thanks to the increasingly available computing power, and medical image segmentation is one of the areas these applications are commonly used. 	\\\\

It is known that the success of the deep learning models heavily relies on the input data volume. In supervised techniques, we also need the annotations of this input data. Getting the necessary annotated data is always costly. Moreover, when we are dealing with  medical problems, a high level of expert knowledge is also required. For this reason, the lack of annotated data proves to be a much more challenging problem  in medical image analysis compared with other imaging problems. This problem especially appears when working with retinal vessel images. These images are being used to diagnose several diseases like diabetes, cardiac diseases, migraine, cataract etc. To perform the correct diagnosis, successful segmentation of the vessels is important because obscured details in the fundoscopic images make the decision making process challenging and hard for the medical experts.  \\\\

Performing the segmentation of the vessels from the fundoscopic image in order to let the medical expert to perform better diagnosis is also a challenging task. The edges of the vessels are extremely thin and quite hard to segment and the quality of the input images are also questionable. Additionally, input images might be noisy and key information can end up  becoming very hard to extract in order to detect diseases. A mistake in this process may cause false positive or false negative diagnosis. Finally, it is important to underline that the quality of the input image can be affected by many things including illumination, sensor noises, incorrect angle, type of the filter used in retinal fundus cameras, etc. \\\\

Successful segmentation of the retinal vessel segmentation has been widely studied and it is still identified as one of the hot topics for research. Different architectures are tailored for this specific problem and numerous existing deep learning architectures are used in order to perform segmentation tasks. The scarcity of the annotated data pushed researchers to use data augmentation to a certain amount in order to avoid the overfitting problem. However, the usage of data augmentation is limited in these studies. This study argues that if data augmentation strategy can successfully address the problems of the input data, a successful segmentation model can be obtained. In our study, we are looking for the performance gains that can be obtained by the excessive data augmentation using U-Net architecture for retinal vessel segmentation problems. We use the DRIVE dataset\footnote{\label{drive}\href{https://drive.grand-challenge.org/}{\texttt{https://drive.grand-challenge.org/}}} that has become one of the standard benchmarks in the retinal vessel segmentation studies.\footnote{All models and methods are available at \href{https://github.com/onurboyar/Retinal-Vessel-Segmentation}{ \texttt{https://github.com/onurboyar/Retinal-Vessel-Segmentation}}}

\begin{figure}%[H]
\centering
\includegraphics[width=9cm]{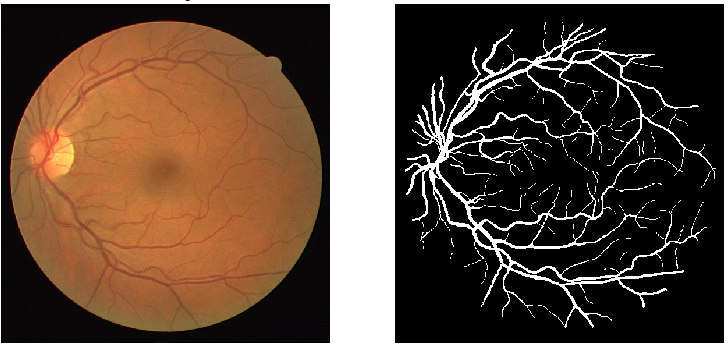}
\caption{Training sample from DRIVE dataset.}
\label{fig:driwe}
\end{figure}

\section{Related Work}
\label{sec:relatedwork}
The Retinal Vessel Segmentation problem has gained much interest in the literature in the last few  years. There are different types of segmentation strategies with varying complexities.
\\\\

Medical Image Segmentation studies rely heavily on the U-Net architecture \cite{ronneberger2015unet}. In \cite{8803101}, MResU-Net is derived from U-Net architecture by replacing convolutional layers with residual blocks in order to achieve better accuracy and increase the depth of the model to infer more features.  In \cite{kamran2021rvgan}, authors introduced a GAN \cite{goodfellow2014generative} based approach to segment the retinal vessels. They take the segmentation problem as an image translation problem. RV-GAN architecture has reported the best accuracy and AUC metrics so far. In \cite{8759448},  a cross connected Convolutional Neural Network architecture is used to perform the segmentation. In \cite{article}, authors proposed an U-Net based model which is combined with a Residual U-Net. They argue  that their proposed model is a good trade-off between the model accuracy and the training time. In \cite{Jin_2019}, authors have also proposed a U-Net based model that is using Deformable Convolutional Networks, DU-Net. In order to train the model, they crop the input images into small batches. Cropping the input image into small batches is a strategy to overcome the problem of scarcity of the annotated training data.\\\\

Data augmentation techniques have been widely used in various problems. Not only in medical imaging segmentation, but also in a wide variety of computer vision problems and in problems which rely on the tabular data \cite{article3}. For the purposes of this study, we limit our review in their usage in image segmentation problems. In \cite{ronneberger2015unet}, the U-Net architecture is proposed for the first time and its usage with data augmentation is also covered. In \cite{EatonRosen2018ImprovingDA}, authors investigate the performance gain as a result of  using data augmentation in medical image segmentation and discuss the common problems of medical image segmentation due to the scarcity of the available annotated data. A comprehensive study of the data augmentation in deep learning models is reported in \cite{Shorten2019ASO}. Data Augmentation in retinal image segmentation problem is not a widely studied area and our study is one of the most comprehensive studies that explores the performance gains from data augmentation in retinal vessel segmentation. In \cite{10.1145/3348416.3348425}, authors lay out the effect of data augmentation in retinal image segmentation. However, their study is limited to rotated augmentations. In \cite{sun2020robust}, the authors focus on looking into  data augmentation in the retinal image segmentation problem and a method that gives a robust segmentation model is proposed. \cite{10.3389/fncom.2019.00083} studies the data augmentation techniques for brain tumor segmentation problem.\\\\

Other than elastic transformations and affine image transformations which are rotation, flipping, scaling and cropping etc., they also study the generative techniques in order to perform the data augmentation. Generative Adversarial Networks have been widely used in medical image segmentation problems for both data augmentation techniques and as a segmentation model itself \cite{article2, 10.3389/fncom.2019.00083}.
\begin{table}
\caption{List of augmentations we use. \label{augmentations}}
\centering
\begin{tabular}{|l|l|l|} 
\hline
\textbf{Affine Transformation} & \textbf{Elastic Transformation} & \textbf{Pixel-level Transformation}  \\ 
\hline
Rotation                       & Elastic Deformations            & White Noise                          \\ 
\hline
Flipping                       & Grid Distortion                 & Gamma Correction                     \\ 
\hline
Zoom Out                       & Optical Distortion              & Equalize Histogram                   \\ 
\hline
Random Cropping                &                                 & Dropout                              \\ 
\hline
Shifting                       &                                 & Sharpening                           \\ 
\hline
Shearing                       &                                 & Blurring                             \\ 
\hline
                               &                                 & Contrast                             \\
\hline
\end{tabular}
\end{table}
\section{Dataset}
In Retinal Vessel Segmentation problem, the most widely used dataset is DRIVE\footnoteref{drive} dataset. Each paper in recent years have reported their model performance metrics on this dataset. DRIVE dataset has 20 train and 20 test images. Each image in the training set addresses a different kind of disease. It is quite challenging to create a high-performance model with this amount of annotated data. We use DRIVE dataset to address these problems.
\section{Experimental Setup}
\subsection{Problem Definition}
The problem we are focusing on is the segmentation of retinal images with input data that has quality problems. Moreover, the amount of the annotated data is very limited. Problems about the input data may occur due to various reasons such as illumination, sensor noise, filters of the retinal camera, the input image angle, and other noises. Using the data with such defects limits the capability of the segmentation model. Most of the time the model is unable to segment the regions with noise due to the fact that the model does not have enough information about the regions trying to be segmented. For example, in the DRIVE dataset, each image in the training set addresses a different disease. Each image may come with different quality problems. Combining this with the input image scarcity problem of the medical imaging problems, the problem at the hand becomes even more difficult. If one can identify the problems of the input images well, the quality of the segmentation model can be increased with the help of the data augmentation.

\subsection{Proposed Methodology}

We propose that optimal data augmentation can be successful for Retinal Vessel Segmentation, using simple U-Net architecture \cite{ronneberger2015unet}. Data augmentation techniques are helpful for three reasons. First, they are helpful because the input data is very scarce. Data augmentation techniques increase the input image size and provide the model some extra information to learn.  Second,  through data augmentation we can recover some performance loss that occurred in the models due to the image quality. If the practitioner decides the data augmentation technique based on the problems of the input image the segmentation performance can be increased. Third,  data augmentation will help with the segmentation model we used, which is the U-Net architecture that makes use of pooling operations. The model learns relatively lower from the corner and side parts of the input image.\\\\

Data augmentation strategies can address all these three problems. In order to address the third problem, we add rotated versions of the input images to the dataset using various angles. We use data augmentation techniques that rely on adding noise to the original image so that our model can learn more from the noisy images. Noise data come from the normal distribution with mean 0 and standard deviation $\epsilon$. In our study we use augmentations with different epsilon values each greater than or equal to 1. Another technique we use is dropout, which targets input pixels. In dropout data augmentation technique, pixels of the input image are set to zero in a random fashion. The portion of the pixel values to be set to zero is a parameter that needs to be defined. It is well-known that minor vessels are one of the hardest regions to segment. Figure 2 shows the predicted and ground truth vessels together. It can be seen that the majority of the incorrect segmentations occur at the minor vessels. An attempt to increase the success of the segmentation model might be to zoom to these regions and add zoomed images to the dataset. Randomly cropping the input image with the random sizes might be another strategy here. We use shifting and flipping of the input image which are also two successful data augmentation techniques. They are widely used with U-Net model training because U-Net uses convolutional filters. Convolutional filters miss the information in the edge of images. Shifting technique pushes the edges of images to the more central part of the image so that the U-Net model can learn the information in the edges of the original image from this augmented image. In order to address the image quality problems that occurred due to the brightness of the input image, we use gamma correction technique. Full list of augmentations are shown in Table 1.

\begin{figure}%[H]
    \centering
    \subfloat[]{{\includegraphics[width=7.5cm]{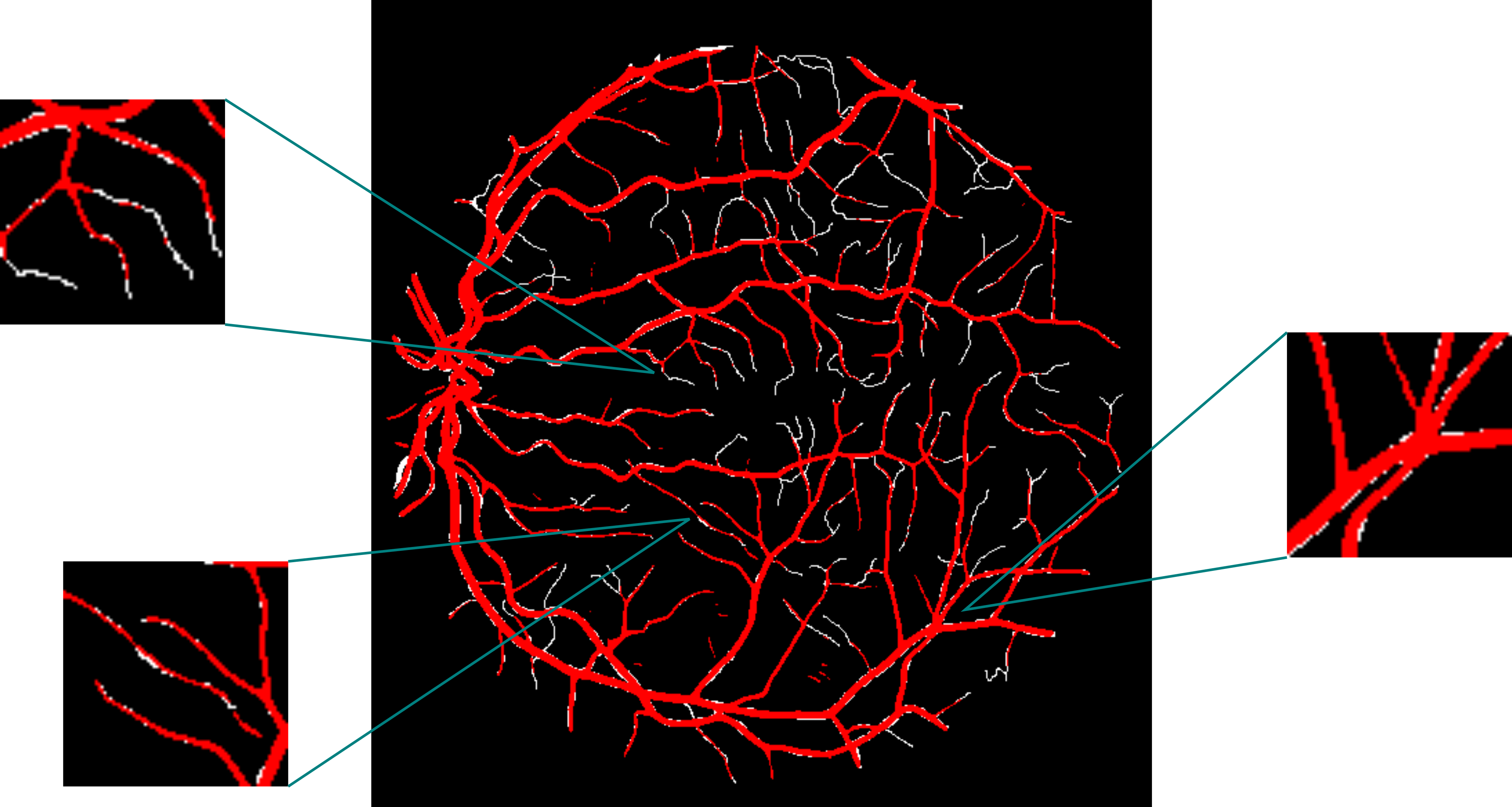} }}%
    \qquad
    \subfloat[]{{\includegraphics[width=7.5cm]{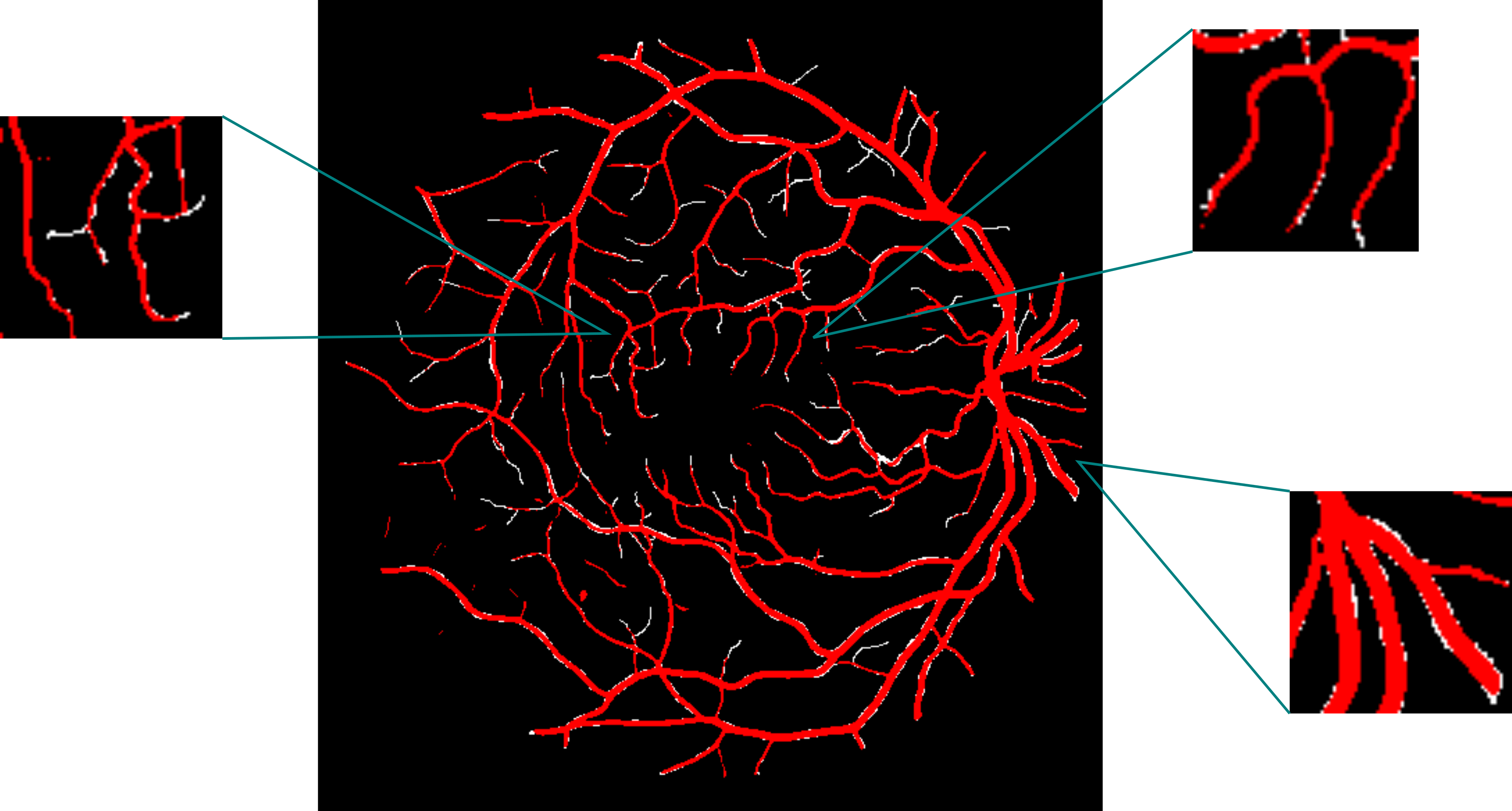} }}%
    \caption{Two examples for a combination of predictions and ground truths. Red pixels are predicted vessels, white pixels are ground truth pixels. By (a), it is observed that the model errors due to the minor vessels. The model performs well in segmenting thick vessels as it is pointed in the right hand side of (a). The model performs better at segmenting the vessels in (b). Minor vessels are segmented better than (a) in this example.}%
    \label{fig:twodriwe}
\end{figure}
\subsection{Implementation Details}
For our method, we use the same architecture from U-Net \cite{ronneberger2015unet}. We train our models with the Adam optimizer \cite{kingma2017adam} with learning rate of 1e-4, $\beta_1 = 0.9$, $\beta_2 = 0.999$. We don’t use any learning rate scheduling algorithms. For experiments on DRIVE, we use mini-batches of size 3. We use a dropout probability of 0.1 on the fourth and fifth convolutional layers. The training is done by using binary cross-entropy loss. In our study, we experimented with dice loss and the combination of binary cross-entropy loss and dice loss \cite{Sudre_2017} as well. Nevertheless, it is observed that the best results are obtained using binary cross-entropy loss. \\\\

We implement all our models in Keras \cite{chollet2015} and train them on a single RTX 2080. We train our models for 15 epochs, each epoch takes 10-15 minutes. We don't use pre-trained weights in our experiments.

\section{Results}
\begin{table}%[ht]
\caption{Performance comparison on the DRIVE dataset.} % title of Table
\centering % used for centering table
\begin{tabular}{c c c c} % centered columns (5 columns)
\hline\hline %inserts double horizontal lines
Paper & Year & AUC\ & Accuracy\ \\ [0.5ex] % inserts table
%heading
\hline % inserts single horizontal line
U-Net \cite{Jin_2019} & 2018 & 0.9830 & 0.9681\\ % inserting body of the table
%DUNet \cite{Jin_2019}  & 2019&0.9856 &    0.9697\\
Residual UNet \cite{8803101} & 2019 & 0.9779 & - \\
IterNet \cite{li2019iternet} & 2019 & 0.9816 & 0.9574  \\
Wang et al. \cite{10.1145/3348416.3348425} & 2019 & 0.9814 & 0.9573 \\
Sun et al. \cite{sun2020robust} & 2020 & 0.9788 & 0.9545\\
CcNet \cite{FENG2020268} & 2020 & 0.9678 & 0.9528 \\
SUD-GAN \cite{sudgan} & 2020 & 0.9786 & 0.9560 \\
RV-GAN \cite{kamran2021rvgan} & 2020 &\textbf{0.9887} & \textbf{0.9790}\\
This Paper & 2021 & \textbf{0.9855} & \textbf{0.9712}  \\ [1ex] % [1ex] adds vertical space
\hline %inserts single line
\end{tabular}
\label{table:nonlin} % is used to refer this table in the text
\end{table}
As seen in Table 2., our approach achieves a 0.9855 Area Under the Curve (AUC) and 0.9712 accuracy score. It outperforms most of the models that have more complex architectures in the literature. Our model also outperforms other data augmentation-based studies in the literature [16, 17]. It is observed that RV-GAN [6] has the highest AUC and accuracy scores. However, the training time of their model takes up to 48 hours. Even though we train our model for about 3 hours, we are only 0.0032 AUC, 0.0078 accuracy points behind RV-GAN. Studies reported in Table 2 did not include the dice coefficient score of their models although it is a common metric in image segmentation tasks. Our method achieved a 0.8255 mean dice score on the DRIVE dataset.

The development of the model with the highest metrics is done by adding new augmentation techniques in each iteration. Different augmentation techniques provided the model a new information and therefore a performance gain. The first set of augmentation techniques that are included are augmentations obtained by rotating and flipping the input images. At each iteration, results of the model have evaluated and new augmentation techniques are added based on the areas that the model failed to perform well. The first observation was the lack of performance in thin vessels which lead us to include shifted, zoomed and cropped images to the model dataset. After gaining considerable amount of accuracy, in the next iteration the impact of white noise and elastic deformations is studied to perform better segmenting the noisy images. Another major problem about the DRIVE dataset is the brightness of some of the input images. To overcome this problem, new augmented images using gamma correction technique are added to model dataset. Finally, extra gain in performance is pursued by adding new augmented images using augmentation techniques like blurring, droput, histogram equalization and distortion techniques.

\section{Conclusion}
In this paper, we propose a segmentation model that relies on the heavy usage of data augmentation techniques. We use the standard U-Net architecture and outperform most of the more complex architectures in the literature. The data augmentation strategy is governed by the problems about the input images caused by the fundus camera and the environment. Various types of data augmentation techniques are used individually and collectively. Data augmentation strategy also takes the drawbacks of U-Net architecture and makes use of various rotated versions of the augmented images into account in order not to lose valuable information due to pooling operations. 

\section{Acknowledgements}
This study is a part of the inzva AI Projects program. We would like to thank to inzva community for providing us a valuable research environment.

\printbibliography %% biblatex with biber backend

% bib compiler natbib is defined as: 
%   - we are using biber backend with biblatex
%\bibliographystyle{unsrtnat}
%\bibliography{references} 





% \end{thebibliography}

\end{document}